\documentclass[pra,twocolumn,showpacs]{revtex4}
\usepackage{graphicx}
\begin{document}

\bibliographystyle{unsrt}

%%%%%%%%%%%%%%%%%%%%%%%%%%%%%%%%%%%%%%%%%
\title{Probabilistic Quantum Encoder for Single-Photon Qubits}
\author{T.B. Pittman, B.C Jacobs, and J.D. Franson}
\affiliation{Johns Hopkins University,
Applied Physics Laboratory, Laurel, MD 20723}
%%%%%%%%%%%%%%%%%%%%%%%%%%%%%%%%%%%%%%%%%

\date{\today}

%%%%%%%%%%%%%%%%%%%%%%%%%%%%%%%%%%%%%%%%%
\begin{abstract}
We describe an experiment in which a physical qubit represented by the  polarization state of a single-photon was probabilistically encoded in the logical state of two photons. The experiment relied on linear optics, post-selection, and three-photon interference effects produced by a parametric down-conversion photon pair and a weak coherent state.  An interesting consequence of the encoding operation was the ability to observe entangled three-photon Greenberger-Horne-Zeilinger states.  
\end{abstract}
%%%%%%%%%%%%%%%%%%%%%%%%%%%%%%%%%%%%%%%%%


\maketitle

One method for protecting quantum information from the effects of noise is the use of redundancy \cite{nielsenchuangbook}. Quantum error correction codes \cite{shor95} are based on the encoding of a single logical qubit in the quantum states of multiple physical qubits.  A simple example of such an encoding is the transformation:

\begin{equation}
\alpha|0\rangle + \beta|1\rangle  \rightarrow  \alpha|00...0\rangle + \beta|11...1\rangle 
\label{eq:encoding}
\end{equation}

\noindent Although full-scale quantum computers will require more complex encoding involving large numbers of physical qubits, small-scale quantum devices may still benefit from simple encoding operations involving a relatively small number of physical qubits.  In this paper, we describe a proof-of-principle experiment in which a single-photon qubit was encoded into a logical state consisting of two photons, as in Eq. (\ref{eq:encoding}).  Operations of this kind are expected to have applications in linear optics quantum computing \cite{knill01a} and quantum communications \cite{jacobs02}.

\begin{figure}[b]
\hspace*{-.25in}
\includegraphics[angle=-90,width=3.75in]{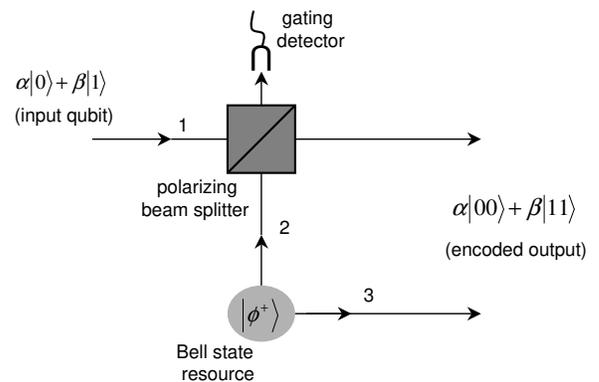}
\vspace*{-.75in}
\caption{Overview of a probabilistic quantum encoding device \protect\cite{pittman01a} in which a polarization-based single-photon input qubit is encoded in the logical state of two output photons: $\alpha|0\rangle + \beta|1\rangle  \rightarrow  \alpha|00\rangle + \beta|11\rangle$. Under idealized conditions, the probability of success of the encoding operation is $\frac{1}{2}$.  }
\label{fig:overview}
\end{figure}

An overview of the quantum encoding device considered here is shown in 
Fig.\ref{fig:overview}.  We originally proposed this encoder as part of a probabilistic controlled-NOT gate, and the complete theory of its operation can be found in reference \cite{pittman01a}.  To briefly review,  the qubits are represented by the polarization states of single photons, with  horizontal and vertical polarization states corresponding to the qubit values $|0\rangle$ and $|1\rangle$. As shown in Fig. \ref{fig:overview}, the encoder consists primarily  of a polarizing beam splitter and a resource pair of entangled photons in the Bell state $|\phi^{+}\rangle = \frac{1}{\sqrt{2}}(|00\rangle +|11\rangle)$.  Consequently, a demonstration of the quantum encoder was essentially a three-photon interference experiment, and the experimental techniques used here were closely related to those used in the demonstration of quantum teleportation \cite{bouwmeester97} as well as, for example, recent work towards quantum repeaters \cite{zhao03,marcikic03}.

For the quantum encoder shown in Fig. \ref{fig:overview}, the input qubit (eg. a single photon in a general polarization state $\alpha|0\rangle + \beta|1\rangle$) and one member of the entangled resource pair are mixed at the polarizing beam splitter. In the idealized case, the eventual detection of exactly one photon by the gating detector signals the fact that the two remaining photons are exiting the device.  Because the beam splitter transmits horizontally polarized photons and reflects vertically polarized photons, it can be shown that the output state is of the form \cite{pittman01a}:

\begin{equation}
|\psi\rangle_{out}=\frac{1}{\sqrt{2}}(\alpha|000\rangle + \beta|111\rangle)+ \frac{1}{\sqrt{2}}|\psi_{\perp}\rangle
\label{eq:encoderoutput}
\end{equation}

\noindent where $|\psi_{\perp}\rangle$ represents a normalized combination of amplitudes that are orthogonal to the condition of finding exactly one photon in the gating detector mode.  The encoding is completed by accepting the output only when the gating detector measures exactly one photon in a polarization basis rotated by $45^{o}$ from the computational basis, and utilizing feed-forward control techniques \cite{pittman02a} to compensate for any resulting phase shifts in the output modes \cite{pittman01a}.  Under these circumstances, which occur with a probability of $\frac{1}{2}$, the device realizes the encoding $\alpha|0\rangle + \beta|1\rangle  \rightarrow  \alpha|00\rangle + \beta|11\rangle$.

%%%%%%%%%%%%%%%%%%%%%%%%%%%%%%%%%%
%%%%%%%%%%%%%%%%%%%%%%%%%%%%%%%%%%%

Our experimental demonstration of this quantum encoding operation suffered from several significant limitations compared to the idealized version shown in Fig. \ref{fig:overview}.  First, the idealized encoder requires an ``on-demand'' (or at least heralded) resource pair of entangled photons (see, for example, \cite{pittman03b} and references therein), whereas our experiment only utilized a random source of entangled pairs produced by parametric down-conversion.  Second, our experiment utilized a gating detector with a limited quantum efficiency, and no ability to distinguish the presence of one versus two photons in the gating mode \cite{fitch03}. Due to these limitations, the operation of the quantum encoder could only be observed in the so-called ``coincidence basis'', in which events were only accepted when the gating detector, as well as additional detectors placed in each of the two device output modes, simultaneously fired. This post-selected mode of operation helped to screen out amplitudes associated with $|\psi_{\perp}\rangle$, as well as overcome the technical problem of photon loss in the apparatus.

\begin{figure}
\includegraphics[angle=-90,width=3.5in]{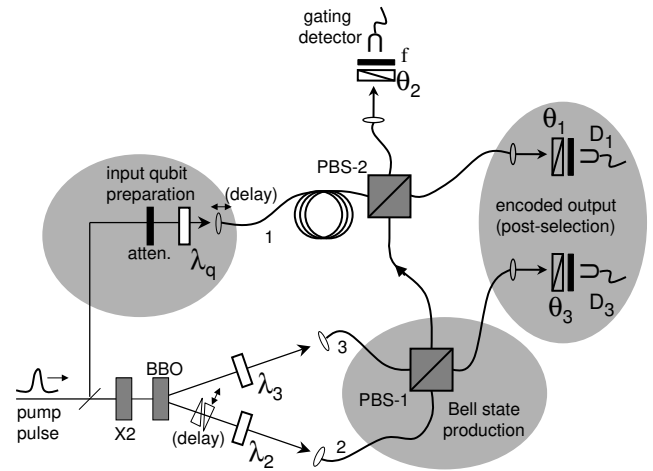}
\caption{Experimental apparatus used to demonstrate the probabilistic quantum encoding operation.  The Bell-state resource pair was derived from a Shih-Alley parametric down-conversion source \protect\cite{shih88}, while the input qubit photon was derived from a weak coherent state.  Additional details are found in the text.}
\label{fig:experiment}
\end{figure}

An outline of the experimental apparatus \cite{pittman03c} is shown in Fig. \ref{fig:experiment}.  It can essentially be thought of as a polarization-based multi-photon interferometer allowing three-photon interference effects among two photons of a parametric down-conversion pair, and a single-photon post-selected from a weak coherent state \cite{rarity97,pittman03a}.  The fidelity of the quantum encoding operation was directly related to the visibility of these three-photon interference effects.  High visibility was primarily achieved by using single-mode fibers to enhance spatial mode-matching of the three separate beams, and pulsed-pump down-conversion with narrowband spectral filters to enhance the temporal overlap of the interfering three-photon amplitudes \cite{zukowski95}.

The entire experiment was driven by a mode-locked Ti-Sapphire laser providing short pulses ($\sim$150fs, 780nm) at a repetition rate of 76MHz.  As shown in Fig. \ref{fig:experiment}, the entangled resource pair of photons was derived from a parametric down-conversion source pumped by frequency doubled (x2, 390nm) laser pulses, while the input qubit photon was provided by weak coherent state pulses picked off from the original laser beam and attenuated to the single-photon level.

The down-conversion crystal (BBO, 0.7mm thick) was optimized for type-I non-collinear down-conversion, producing pairs of horizontally polarized photons at 780nm.   Half-wave plates ($\lambda_{2}$ and $\lambda_{3}$) were used to rotate the polarizations of these photons to $45^{o}$, and the photons were subsequently coupled into the single-mode fiber input ports (labelled 2 and 3) of polarizing beam splitter PBS-1.  A delay unit formed by translating glass wedges was adjusted so that the down-converted photons arrived at PBS-1 well within their coherence times \cite{hong87}. For a given input pair, there was a 50\% chance that one photon would exit each port of PBS-1, in which case the resource Bell state $|\phi^{+}\rangle = \frac{1}{\sqrt{2}}(|00\rangle +|11\rangle)$ was produced.  Those cases in which both photons of a down-conversion pair exited the same port of PBS-1 would not contribute to the coincidence-basis three-photon events of interest, aside from a small noise term which will be described below.  This method of post-selecting the resource Bell-state $|\phi^{+}\rangle$ was equivalent to the technique used in the original Shih-Alley Bell's inequalities tests \cite{shih88}. The fidelity of the post-selected two-photon Bell states produced and measured in our setup was typically about 97\%.

%%%%%%%%%%%%%%%%%%%%%%%%%%%%%%%%%%
%%%%%%%%%%%%%%%%%%%%%%%%%%%%%%%%%%%

The input qubit photon was coupled into one fiber input port (labelled 1) of PBS-2, which was analogous to the main polarizing beam splitter of the idealized quantum encoder illustrated in Fig. \ref{fig:overview}.  The input qubit's logical value could be specified by rotating the polarization state with half-wave plate $\lambda_{q}$. A variable delay in the input qubit photon's path was used to ensure that the input qubit and the relevant member of the Bell state resource pair arrived at PBS-2 within their coherence times \cite{rarity97,pittman03a}.

The gating detector was preceded by a polarization analyzer ($\theta_{2}$) fixed at $45^{o}$, as required by the measurement condition of eq.(\ref{eq:encoderoutput}).  The additional detectors in the output modes of the quantum encoder ($D_{1}$ and $D_{3}$) were preceded by rotating analyzers that were used to measure and verify the polarization state of the encoded output for various examples of input qubit values.  All three detectors were preceded by interference filters (f) centered at 780nm with a bandwidth of 10nm \cite{zukowski95}.

The three-fold coincident events of interest corresponded to the detection of the two photons of the Bell state resource pair, and a single photon from the weak coherent state input qubit pulse.  
The largest source of noise in the three-fold counting rate corresponded to those cases in which one (or two) down-converted photon(s) triggered $D_{3}$ while events at the other two detectors were due to the small probability that the weak coherent state pulse actually contained two photons. Consequently, there was a trade-off between increasing the magnitude of the weak coherent state in an effort to increase the overall valid three-photon detection rate, while keeping it low enough to maintain the error contribution at an insignificant level.  

The optimal value for the magnitude of the weak coherent state was dictated by the percentage of detections at $D_{3}$ that resulted in a detection of a second down-conversion photon at one of the other two detectors.
In our setup this value was measured to be roughly 10\%, which was primarily due to inefficient coupling of the down converted photons into fibers 2 and 3, limited detection efficiency, and loss in the various optical components and connectors \cite{typeiivsi}.  The magnitude of the weak coherent state was therefore adjusted so that the probability per pulse of detecting a single input qubit photon was on the order of $10^{-3}$, which kept the ratio of error events to valid three-photon events on the order of $10^{-2}$.

\begin{figure}[b]
\hspace*{-.15in}
\includegraphics[angle=-90,width=3.65in]{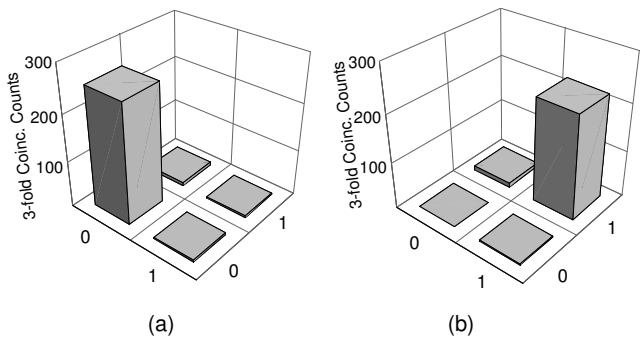}
\vspace*{-1in}
\caption{Experimental results demonstrating the basis-state encoding operations $|0\rangle \rightarrow |00\rangle$ (a), and $|1\rangle \rightarrow |11\rangle$ (b).  The data shows the number of three-fold detection events per 1200 seconds for polarization analyzer settings of $|\theta_{1}$,$\theta_{3}\rangle$ corresponding to output states $|00\rangle$,$|01\rangle$$,|10\rangle$, and $|11\rangle$.}
\label{fig:basisstates}
\end{figure}

As a demonstration of the probabilistic quantum encoder, we accumulated data for several different logical values of the input qubit.  Fig. \ref{fig:basisstates} shows the results obtained with input qubit having the value of either $|0\rangle$ or $|1\rangle$.  The data clearly shows the expected encoding operations, $|0\rangle \rightarrow  |00\rangle$ and $|1\rangle \rightarrow |11\rangle$. The results of these basis-state examples only relied upon the reflection and transmission properties of the beam splitters, and did not critically depend on three-photon interference effects.  As a result, the relatively small errors in these results were primarily due to minor beam splitter imperfections, imperfectly compensated birefringence in the optical fibers, and asymmetric losses.

%%%%%%%%%%%%%%%%%%%%%%%%%%%%%%%%%%
%%%%%%%%%%%%%%%%%%%%%%%%%%%%%%%%%%%

\begin{figure}
\hspace*{-.1in}
\includegraphics[angle=-90,width=3.55in]{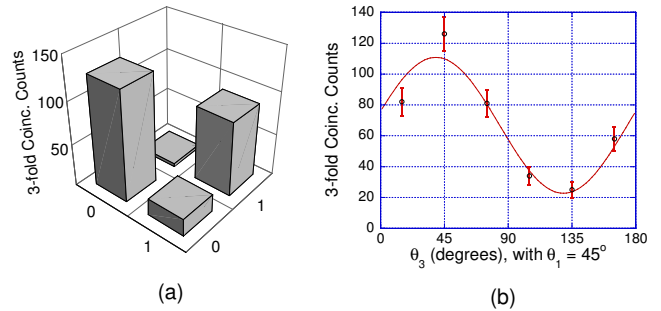}
\vspace*{-1in}
\caption{Experimental results obtained for the input qubit value $\frac{1}{\sqrt{2}}(|0\rangle + |1\rangle)$ (eg. $45^{o}$), in which case the encoder is expected to produce the entangled state $\frac{1}{\sqrt{2}}(|00\rangle + |11\rangle)$ in its output.  (a) shows results demonstrating the existence of the $|00\rangle$ and $|11\rangle$ terms, and the suppression of the unwanted $|01\rangle$ and $|10\rangle$ terms due to destructive three-photon interference effects. (b) demonstrates the coherence between the $|00\rangle$ and $|11\rangle$ terms. The data shows the number of three-fold events as a function of $\theta_{3}$, with the $\theta_{1}$ setting fixed at the logical value $\frac{1}{\sqrt{2}}(|0\rangle + |1\rangle)$ (eg. $45^{o}$). The solid line is a sinusoidal fit to the data, with a visibility of (66.0 $\pm$ 5.8 \%).  In both plots, the data accumulation time was 1200 seconds per point.}
\label{fig:bellstate}
\end{figure}

The data shown in Fig. \ref{fig:bellstate} corresponds to the case when the input qubit was chosen to be in the symmetric superposition state $\frac{1}{\sqrt{2}}(|0\rangle + |1\rangle)$ (eg. $45^{o}$), in which case the encoder is expected to produce the $\phi^{+}$ Bell state $\frac{1}{\sqrt{2}}(|00\rangle + |11\rangle)$ in its output. In addition to the technical error sources mentioned above, the vanishing of the unwanted $|01\rangle$ and $|10\rangle$ cross-terms in this case did critically depend on destructive three-photon interference effects.   In our experiment the visibility of these interference effects was typically in the range of 65 - 70\% , and was primarily limited by the use of relatively wide (10nm) bandwidth interference filters \cite{pittman03a,zukowski95} to allow larger three-photon counting rates.  Consequently, although the data shown in Fig. \ref{fig:bellstate} does indicate that the encoded output was in the $\phi^{+}$ Bell state, the results were insufficient for an explicit violation of Bell's inequalities \cite{clauser78}.

%%%%%%%%%%%%%%%%%%%%%%%%%%%%%%%%%%
%%%%%%%%%%%%%%%%%%%%%%%%%%%%%%%%%%%
 
Finally, an interesting feature of our experimental apparatus was the ability to observe post-selected three-photon Greenberger-Horne-Zeilinger (GHZ) correlations \cite{greenberger90}.  It can be seen from equation (\ref{eq:encoderoutput}) that if the input qubit is in a symmetric superposition state (eg. $\alpha=\beta$), then the detection of exactly one photon by each of the three detectors would post-select the entangled state $|\psi_{GHZ}\rangle =\frac{1}{\sqrt{2}}(|000\rangle+|111\rangle)$. 
This represents an experimental realization of a polarization-based three-photon entanglement scheme that was first proposed by Rarity and Tapster \cite{rarity99}.  In contrast to the original observations of GHZ correlations that used two down-conversion photon pairs (see, for example \cite{bouwmeester99}), the basic idea here is to produce a three-photon entangled state using two photons from a single down-conversion pair, and a third photon from a weak coherent state.

Whereas the operation of the quantum encoder required the analyzer in front of the gating detector to be fixed at $45^{o}$, the GHZ state of interest here could be revealed by accumulating data at multiple settings of all three analyzers.  As a first test, we gathered data at the eight combinations of basis-state analyzer settings needed to verify the existence of only the $|000\rangle$ and $|111\rangle$ amplitudes, and the suppression of the six unwanted cross terms ($|001\rangle$, $|010\rangle$...etc.). The measured ratio of desired to undesired events in this case was roughly 19:1.  

As a second test, data gathered with the analyzers set to $\pm 45^{o}$ could be used to provide evidence that the $|000\rangle$ and $|111\rangle$ amplitudes were indeed in the coherent superposition state $|\psi_{GHZ}\rangle$, rather than some statistical mixture of these two terms. An example of these kinds of measurements can be inferred from the data shown in Figure \ref{fig:bellstate}.  Once again, the incomplete cancellation of the cross terms in Figure \ref{fig:bellstate}(a), and the 66\% visibility (rather than 100\% visibility) of the three-photon interference pattern in Figure \ref{fig:bellstate}(b) indicates some contamination of the ideal $|\psi_{GHZ}\rangle$ state due to experimental inefficiencies.  Nonetheless, experiments of this kind could be used in tests of quantum nonlocality through Bell-type inequalities for more than two particles (see, for example, \cite{mermin90}). 
 
In summary, we have demonstrated a quantum encoder for single-photon qubits.  Devices of this kind are expected to have applications in linear optics quantum gates \cite{pittman01a} and various quantum communications protocols \cite{jacobs02}.  For many of these applications, an efficient photon number resolving detector \cite{fitch03} and a heralded source of entangled photons \cite{pittman03b} will be required.  Although the results presented here were obtained in the coincidence basis instead, they still demonstrate the basic features of a quantum encoder.  In addition, these three-photon interference effects also demonstrated a new source of polarization-based GHZ states in which two photons were obtained from parametric down-conversion while the third photon was post-selected from a weak coherent state \cite{rarity99}.

This work was supported by ARO, NSA, ARDA, ONR, and IR\&D funding.

%%%%%%%%%%%%%%%%%%%%%%%%%%%%%%%

%%%%%%%%%%%%%%%%%%%%%%%%%%%%%%%%


\begin{thebibliography}{50}

\bibitem{nielsenchuangbook} {\em Quantum Computing and Quantum Information}, M.A.
Nielsen and I.L. Chuang, Cambridge University Press (2000).

\bibitem{shor95}  P.W. Shor, Phys. Rev. A {\bf 52}, 2493 (1995).

\bibitem{knill01a} E. Knill, R. Laflamme, and G.J. Milburn, Nature {\bf 409}, 46
(2001).

\bibitem{jacobs02} B.C. Jacobs, T.B. Pittman, and J.D. Franson,  Phys. Rev. A {\bf 66}, 052307 (2002).

\bibitem{pittman01a} T.B. Pittman, B.C. Jacobs, and J.D. Franson, Phys. Rev. A {\bf 64}, 062311 (2001).

\bibitem{pittman02a} T.B. Pittman, B.C. Jacobs, and J.D. Franson, Phys. Rev. A {\bf 66}, 052305 (2002).

\bibitem{bennett93} C.H. Bennett {\em et.al.}, Phys. Rev. Lett. {\bf 70}, 1895 (1993).

\bibitem{bouwmeester97} D. Bouwmeester {\em et.al.}, Nature {\bf 390}, 575 (1997); J.-W. Pan {\em et.al.}, Nature {\bf 421}, 721 (2003).

\bibitem{zhao03} Z. Zhao, T. Yang, Y.-A. Chen, A.-N. Zhang, and J.-W. Pan, Phys. Rev. Lett. {\bf 90}, 207901 (2003).

\bibitem{marcikic03} I. Marcikic, H. de Riedmatten, W. Tittel, H. Zbinden, and N. Gisin, Nature {\bf 421}, 509 (2003).

\bibitem{pittman03b} T.B. Pittman {\em et.al.}, IEEE J. Select Topics Quant. Elec. (2003); quant-ph/0313113.

\bibitem{fitch03}  M.J. Fitch, B.C. Jacobs, T.B. Pittman, and J.D. Franson, Phys. Rev. A {\bf 68}, 043814 (2003).

\bibitem{pittman03c} T.B. Pittman, M.J. Fitch, B.C. Jacobs, and J.D. Franson, Phys. Rev. A {\bf 68}, 032316 (2003).

\bibitem{rarity97} J.G. Rarity and P.R. Tapster, Philos. Trans. R. Soc. London {\bf A 355}, 2267 (1997); quant-ph/9702032.

\bibitem{pittman03a} T.B. Pittman and J.D. Franson, Phys. Rev. Lett. {\bf 90}, 240401 (2003).

\bibitem{zukowski95} M. Zukowski, A. Zeilinger, and H. Weinfurter, Ann. N.Y. Acad. Sci. {\bf 755} 91, (1995). J.G. Rarity, {\em ibid} {\bf 755}, 624 (1995).

\bibitem{hong87} C.K. Hong, Z.Y. Ou, and L. Mandel, Phys. Rev. Lett. {\bf 59}, 2044 (1987).

\bibitem{shih88} Y.H. Shih and C.O. Alley, Phys. Rev. Lett. {\bf 61}, 2921 (1988).

\bibitem{typeiivsi} In principle, this 10\% value could be increased by a modest factor of two by deriving the Bell state resource pair from a non-collinear type-II down-conversion source (eg. P.G. Kwiat {\em et. al.} Phys. Rev. Lett. {\bf 75}, 4337 (1995))  rather than the Shih-Alley technique \protect\cite{shih88} used here.

\bibitem{clauser78} J.F. Clauser and A. Shimony, Rep. Prog. Phys. {\bf 41}, 1881 (1978).

\bibitem{greenberger90} D.M. Greenberger, M.A. Horne, A. Shimony, and A. Zeilinger, Am. J. Phys {\bf 58}, 1131 (1990).

\bibitem{rarity99} J.G. Rarity and P.R. Tapster, Phys. Rev. A {\bf 59}, R35 (1999).

\bibitem{bouwmeester99} D. Bouwmeester, J.-W. Pan, M. Daniell, H. Weinfurter, and A. Zeilinger, Phys. Rev. Lett. {\bf 82}, 1345 (1999).

\bibitem{mermin90} N.D. Mermin, Phys. Rev. Lett. {\bf 65}, 1838 (1990); A.V. Belinsky and D.N. Klyshko, Phys. Lett. A {\bf 176}, 415 (1993).

\end{thebibliography}
\end{document}